\documentclass[12pt]{article}

\usepackage{amsmath}
\usepackage{amssymb}
\usepackage{amsthm}
\usepackage{amsfonts}
\usepackage{mathrsfs}
\usepackage{dsfont}
\usepackage{slashed}
\usepackage{xspace}

\newcommand{\chiPT}{$\chi$PT\xspace}
\newcommand{\tr}{\text{Tr}}

\newcommand{\CLsing}{{}^{1}C_L}
\newcommand{\CRsing}{{}^{1}C_R}
\newcommand{\CLtrip}{{}^{3}C_L}
\newcommand{\CRtrip}{{}^{3}C_R}
\newcommand{\CLRsing}{{}^{1}C_{L/R}}
\newcommand{\CLRtrip}{{}^{3}C_{L/R}}
\newcommand{\tCLtrip}{{}^{3}\widetilde{C}_L}
\newcommand{\tCRtrip}{{}^{3}\widetilde{C}_R}
\newcommand{\tCLRtrip}{{}^{3}\widetilde{C}_{L/R}}

\newcommand{\Id}{\mathds{1}}

\begin{document}

\begin{titlepage}

\begin{flushright}
arXiv:1712.00838
\end{flushright}
\vskip 2.5cm


\begin{center}
{\Large \bf Hadronic Lorentz Violation in Chiral Perturbation\\
Theory Including the Coupling to External Fields}
\end{center}

\vspace{1ex}

\begin{center}
{\large Rasha Kamand\footnote{{\tt kamand@email.sc.edu}},
Brett Altschul\footnote{{\tt baltschu@physics.sc.edu}},
and Matthias R.~Schindler\footnote{\tt {mschindl@mailbox.sc.edu}}}

\vspace{5mm}
{\sl Department of Physics and Astronomy} \\
{\sl University of South Carolina} \\
{\sl Columbia, SC 29208} \\

\end{center}

\vspace{2.5ex}

\medskip

\centerline {\bf Abstract}

\bigskip

If any violation of Lorentz symmetry exists in the hadron sector, its ultimate origins must
lie at the quark level. We continue the analysis of how the theories
at these two levels are connected, using chiral perturbation theory. Considering
a two-flavor quark theory, with dimension-4 operators that break Lorentz
symmetry, we derive a low-energy theory of pions and nucleons that is
invariant under local chiral transformations and includes the coupling to external fields. The pure meson and baryon sectors,
as well as the couplings between them and the couplings to external
electromagnetic and weak gauge fields, contain forms of Lorentz violation which
depend on linear combinations of quark-level coefficients. In particular, at leading
order the electromagnetic couplings depend on the very same combinations as appear in
the free particle propagators. This means that observations of electromagnetic
processes involving hadrons---such as vacuum Cerenkov radiation, which may be
allowed in Lorentz-violating theories---can only reliably constrain certain particular combinations
of quark coefficients.

\bigskip

\end{titlepage}

\newpage

\section{Introduction}
\label{sec:intro}

For the last two decades, there has been a renewed interest in the possibilities
for the seemingly fundamental Lorentz and CPT symmetries to be violated in nature.
There is, as yet, no compelling evidence for such an exotic suggestion.
However, if such symmetry breaking ever were discovered experimentally, the
discovery would obviously be of truly tremendous importance. There are currently
two fundamental theories that explain all the physics we presently understand;
these are the standard model of particle physics (a relativistic quantum field
theory) and general relativity. Local Lorentz invariance is a basic element
of each of these theories, and if the symmetry is found to broken, that could
tell us a great deal about how new physics will differ from the theories we
currently understand.

The reasons for the recent interest in Lorentz violation are several. First, there
is the reason alluded to above; the symmetry is so fundamental that if it does turn
out to be broken, the consequences would be extremely far reaching. Second, there
has been a realization that many of the frameworks that have been suggested as
possibilities for describing quantum gravity seem to allow for the existence of
Lorentz-violating (LV) regimes. Third, studies motivated by the first two
reasons led to the development of a comprehensive effective field theory (EFT)
describing LV physics; and it was found that there was a much richer array of
possible forms that the Lorentz violation could take than had previously been
realized. It was found that there were large regions of the EFT parameter space
that had been constrained only very poorly (or possibly not at all) by earlier
generations of experiments. Furthermore, it turns out that  CPT symmetry is closely
tied to Lorentz symmetry, to the extent that it is not possible to have a CPT-violating
theory with a well-defined $S$-matrix, without it also displaying Lorentz
violation~\cite{ref-greenberg}. (This result holds even if the theory is
allowed to be nonlocal.) As a result, a single EFT suffices to describe both
Lorentz and CPT violations.

The EFT that has been developed to address questions about Lorentz violation is
known as the standard model extension (SME)~\cite{ref-kost1,ref-kost2}. Its action
is built using standard model fields. In the same way that the standard model is
formed by including all local, renormalizable, gauge-invariant, Lorentz-conserving
(LC) operators that can be built out of those fields, the SME is likewise built
up, but without the requirement that the terms of the Lagrange density
be Lorentz scalars.  As an EFT, the SME really has an infinite number of operators,
but if the tower of operators is truncated by introducing the usual conditions of
locality, renormalizability, and $SU(3)_{c}\times SU(2)_{L}\times U(1)_{Y}$
gauge invariance, the result is the minimal SME (mSME). The mSME has become the
default framework for parametrizing the results of most experimental Lorentz and CPT
tests. Since the mSME contains a very large number of independent parameters,
many different kinds of experiments have been useful in placing bounds on those
parameters. A current summary of the results of all these tests is given
in~\cite{ref-tables}.

However, there are weaknesses to the SME formalism as well. One of the most
significant is that the SME is formulated in terms of the fundamental fields of
the standard model. In the strongly interacting sector, this means quarks and
gluons. Of course, these quanta are confined inside hadrons and are not
directly observable. Experimental searches for LV phenomena involving
strongly interacting particles are done with hadrons or nuclei. It is common
practice to translate the results of such experiments into bounds on effective
SME parameters for the hadrons involved---be they protons, neutrons, pions, or
other particles. Bounds are then typically quoted as if the composite hadrons
were really the fundamental excitations of the theory. There are many
extremely tight bounds that have been formulated this way, but it would be
desirable if these could be translated into bounds on the fundamental SME
parameters that apply at the quark and gluon level.

This paper continues our investigation of this last question, and we shall
see that there are some subtleties to the proper analysis, which previous
work has sometimes overlooked.
Using chiral perturbation theory (\chiPT) \cite{Weinberg:1978kz,Gasser:1983yg,Gasser:1984gg} techniques, in~\cite{Kamand:2016xhv} we
described the construction of an effective Lagrangian at the hadronic level
corresponding to the allowed LV modifications of the quark kinetic term in the
mSME. In that paper, the Lagrange density was required to be invariant under {\em global}
chiral $SU(2)_{L}\times SU(2)_{R}$ transformations, and no coupling to external gauge
fields was considered. Here, we shall extend the discussion by promoting the invariance
of the Lagrange density to include {\em local} chiral
transformations~\cite{Gasser:1983yg,Gasser:1984gg} and by including the couplings
of pions and nucleons to external fields.
In addition to the inclusion of external fields, the local invariance of the Lagrangian ensures that the Ward
identities of the underlying theory are satisfied \cite{Leutwyler:1993iq}.
As usual in \chiPT, the identification of the most relevant terms in the Lagrangian will be guided by a power
counting scheme, with operators involving higher powers of the spatial momentum being
successively more suppressed.
Different LV terms have also recently been analyzed in the \chiPT framework \cite{ref-noordmans,ref-noordmans1}.

This paper is organized as follows. In section~\ref{sec:chpt}, we introduce the two-flavor
\chiPT model with local chiral transformations, initially without any Lorentz violation
included. LV terms are then introduced at the quark level in section~\ref{sec:quark-level}, in a
way that maintains the local chiral symmetry. From the structure of the theory at the quark
level, we shall map out the forms taken by the LV operators at the hadronic level in
section~\ref{sec:LV_ChPT}. We shall derive the Lagrangian for mesons first, before turning to
couplings, including the coupling to an external electromagnetic gauge field---a topic which has
significant phenomenological implications. Section~\ref{sec:concl} summarizes our
conclusions and gives some perspective on the experimental significance of our \chiPT
results.

\section{Chiral Perturbation Theory}
\label{sec:chpt}

The starting point of $SU(2)$ \chiPT is the two-flavor quantum chromodynamics (QCD) Lagrange
density in the limit of massless $u$ and $d$ quarks,\footnote{To avoid the excessive use of
super- and subscripts, we use $\mathscr{L}$ for LC and $\mathcal{L}$ for LV Lagrange densities.}
\begin{equation}
\label{masslessQCD}
\mathscr{L}^0_\text{QCD} = \bar{Q}_L i \slashed{D} Q_L + \bar{Q}_R i \slashed{D} Q_R - \frac{1}{2}\text{Tr}(G_{\mu\nu}G^{\mu\nu}).
\end{equation}
Here  $Q_{L/R}=[u_{L/R},d_{L/R}]^T$ denote the doublets of left- and right-handed quark fields;
$D_\mu q = (\partial_\mu + i g A_\mu) q $ is the QCD covariant derivative, with $A_\mu$ the
gluon fields and $g$ the strong coupling constant; and $G_{\mu\nu}$ is the gluon field strength tensor.
In the limit of vanishing up and down quark masses, QCD is invariant under {\em global} chiral
transformations,
\begin{equation}
Q_L \mapsto L Q_L,\quad Q_R \mapsto R Q_R,
\end{equation}
with $(L,R) \in SU(2)_L \times SU(2)_R$. This symmetry is assumed to be spontaneously broken
to the group $SU(2)_V$, with the pions as the associated Goldstone bosons. In the transition to
the hadronic effective Lagrangian, the pions can be collected in the  $SU(2)$ matrix~\cite{Coleman:1969sm}
\begin{equation}
U(x) = \exp\left[ i\frac{\phi(x)}{F} \right] = \exp\left[ i\frac{\sum\phi_i(x)\tau_i}{F} \right],
\end{equation}
where the $\phi_i$ for $i=1,2,3$ are the canonical pion fields, the $\tau_i$ are the Pauli
matrices, and $F$ is the pion decay constant in the chiral limit. Under chiral transformations the pion fields transform as
\begin{equation}
U\mapsto R U L^\dagger .
\end{equation}
The effective Lagrangian describing pions and their interactions is constructed in terms of
$U(x)$ and its derivatives by considering all terms consistent with the symmetries of
QCD---including, in particular, chiral symmetry.
These terms are arranged according to a power counting scheme~\cite{Weinberg:1978kz}, which
assigns a chiral order to each term, with higher chiral orders suppressed at low energies.
In particular, derivatives acting on pion fields correspond to powers of momenta, which are
suppressed at low energies. The leading-order (LO) pion Lagrange density is given by
\begin{equation}
\label{LC_massless_pion_LO}
\mathscr{L}_\pi^\text{LO,0} = \frac{F^2}{4}\tr(\partial_\mu U \partial^\mu U^\dagger).
\end{equation}

In the real world, the light quarks are massive and also interact via the electroweak
interactions. To include these effects, the quark masses and electroweak gauge fields are
treated as external fields in the hadronic effective theory, and the QCD Lagrangian in
the chiral limit is emended with the inclusion of couplings of (axial) vector currents
and (pseudo-)scalar quark densities to $c$-number fields~\cite{Gasser:1983yg,Gasser:1984gg},
\begin{equation}
\label{QCD-external}
\mathscr{L} =  \mathscr{L}^0_\text{QCD} + \mathscr{L}_\text{external}.
\end{equation}
The coupling to the external fields is described by
\begin{align}
\label{LagExtLC}
\mathscr{L}_\text{external}  & = \bar{Q}_L \gamma^\mu \left(l_\mu+\frac{1}{3}v_\mu^{(s)} \right)Q_L
+ \bar{Q}_R \gamma^\mu \left(r_\mu+\frac{1}{3}v_\mu^{(s)} \right)Q_R \nonumber \\
& \quad + \bar{Q}_L (s-ip)Q_R + \bar{Q}_R (s+ip)Q_L ,
\end{align}
where we have introduced the external fields $l_{\mu}$, $r_{\mu}$, $s$, and $p$.
It is sometimes more convenient to consider the traceless vector and axial vector currents given by
\begin{equation}
l_\mu = v_\mu - a_\mu,\quad r_\mu = v_\mu + a_\mu ,
\end{equation}
where
\begin{equation}
v^\mu = \sum_{i=1}^3 \frac{\tau_i}{2} v^\mu_i ,\quad a^\mu = \sum_{i=1}^3 \frac{\tau_i}{2} a^\mu_i .
\end{equation}
The nonzero quark masses are taken into account by setting $s=\mathcal{M} = \text{diag}(m_u,m_d)$. Similarly, making
appropriate substitutions for $l_{\mu}$ and $r_{\mu}$
introduces the couplings of pions to photons and weak gauge bosons. For example, setting
\begin{equation}
\label{ExEMCouple}
l_\mu=r_\mu= -\frac{e}{2} \mathcal{A}_\mu \tau_3,\quad v_\mu^{(s)} = -\frac{e}{2} \mathcal{A}_\mu ,
\end{equation}
gives the coupling to an external electromagnetic four-vector potential $\mathcal{A}_\mu$.
While we shall not consider the singlet axial vector current because of the chiral anomaly, we do need to include the
singlet vector current $v_\mu^{(s)}$, which is in principle related to an additional $U(1)_V$ symmetry.
The reason for including the singlet vector current is that, in the $SU(2)$ case considered here, the quark
charge matrix has a nonzero trace, and thus the coupling to an external electromagnetic field cannot
be described in terms of the traceless matrices $l_\mu$ and $r_\mu$ alone. For this reason, it is
convenient in the mesonic theory to replace the fields $l_\mu$ and $r_\mu$ by
$\tilde l_\mu = l_\mu +\frac{1}{3} v_\mu^{(s)}\mathds{1}$ and
$\tilde r_\mu = r_\mu +\frac{1}{3} v_\mu^{(s)}\mathds{1}$, respectively. (See,
e.g., Ref.~\cite{Ebertshauser:2001nj}.)

Both the light quark masses and the coupling to the electroweak gauge fields break chiral symmetry.
However, the pattern of chiral symmetry breaking can be mapped onto the effective theory by
treating the external fields as spurion fields, which transform in such a way as to maintain
invariance under chiral transformations.
The Lagrange density of eq.~\eqref{QCD-external} is invariant under \emph{local} chiral
$SU(2)_L\times SU(2)_R$ transformations,
\begin{equation}
\label{local_transform}
Q_L \mapsto V_L(x) Q_L , \quad Q_R \mapsto V_R(x) Q_R,
\end{equation}
where $(V_L,V_R)\in SU(2)_L \times SU(2)_R$, provided the external fields transform as
\begin{align}
\label{ext_transform}
l_\mu & \mapsto V_L l_\mu V_L^\dagger + i V_L \partial_\mu V_L^\dagger, \notag \\
r_\mu & \mapsto V_R r_\mu V_R^\dagger + i V_R \partial_\mu V_R^\dagger, \notag \\
v_\mu^{(s)} & \mapsto v_\mu^{(s)}, \\
s+ip & \mapsto V_R (s+ip) V_L^\dagger, \notag \\
s-ip & \mapsto V_L (s-ip) V_R^\dagger . \notag
\end{align}
Since the (constant) quark mass matrix is included simply by setting $s=\mathcal{M}$,
the transformations of eq.~\eqref{ext_transform} imply $\mathcal{M} \mapsto V_R \mathcal{M} V_L^\dagger$
under chiral transformations.\footnote{In this approach, $\mathcal{M}$ is formally
distinguished from $\mathcal{M}^\dagger$, despite the quark masses being real parameters.}

In the hadronic effective theory, the external fields are also used as additional building
blocks in constructing a chirally invariant Lagrange density. To ensure invariance under {\em local}
chiral transformations, the derivative of $U(x)$ should be replaced by a covariant derivative
\begin{equation}
\label{pion_cov}
D_\mu U \equiv \partial_\mu U -i r_\mu U + i U l_\mu,
\end{equation}
which transforms  under local chiral transformations as
\begin{equation}
D_\mu U \mapsto V_R D_\mu U V_L^\dagger.
\end{equation}
It is also convenient to introduce the following building blocks
\begin{align}
\chi &\equiv 2B(s+ip), \nonumber \\
f_R^{\mu\nu} &\equiv \partial^\mu r^\nu - \partial^\nu r^\mu - i [r^\mu, r^\nu],  \\
f_L^{\mu\nu} &\equiv \partial^\mu l^\nu - \partial^\nu l^\mu - i [l^\mu, l^\nu], \nonumber
\end{align}
where $B$ is a constant proportional to the scalar singlet quark condensate, $B=-\frac{1}{2} \langle \bar{Q} Q \rangle$.
In the effective theory, it relates the square of the pion mass to the light quark masses, $M_\pi^2 = B(m_u + m_d)$.
A direct determination of $B$ from QCD requires a nonperturbative calculation. The building blocks transform under
local chiral transformations as
\begin{align}
\chi &\mapsto V_R \chi V_L^\dagger, \nonumber \\
f_R^{\mu\nu} &\mapsto V_R  f_R^{\mu\nu} V_R^\dagger,  \\
f_L^{\mu\nu} &\mapsto V_L f_L^{\mu\nu} V_L^\dagger. \nonumber
\end{align}
In the power counting scheme the building blocks are counted as
\begin{equation}
U= {\cal O}(q^0), \quad D^\mu U = {\cal O}(q), \quad \chi = {\cal O}(q^2), \quad f_{L/R}^{\mu\nu} = {\cal O}(q^2),
\end{equation}
where $q$ is a small expansion parameter.
The LO mesonic Lagrangian, including nonzero quark masses and the couplings to external fields, is given by
\begin{equation}
\label{LC_pion_LO}
\mathscr{L}_\pi^\text{LO} = \frac{F^2}{4}\tr(D_\mu U D^\mu U^\dagger) +
\frac{F^2}{4}\tr(\chi U^\dagger + U \chi^\dagger).
\end{equation}

When considering nucleons, the nucleon doublet $\Psi = [p,n]^T$ transforms
as~\cite{Coleman:1969sm,Callan:1969sn,Georgi:1985kw}
\begin{equation}
\Psi \mapsto K(V_L,V_R,U) \Psi,
\end{equation}
where $K(V_L,V_R,U)$ is defined by
\begin{equation}
\label{KDef}
u(x) \mapsto u^\prime(x) = \sqrt{V_RUV_L^\dagger} \equiv V_R u K^\dagger(V_L,V_R,U) = K(V_L,V_R,U) u V_L^\dagger,
\end{equation}
with $[u(x)]^2=U(x)$. The covariant derivative acting on the nucleon fields is given by
\begin{equation}
\label{NucCovDer}
D_\mu \Psi \equiv \left[\partial_\mu + \Gamma_\mu - i v_\mu^{(s)} \right]\Psi,
\end{equation}
where
\begin{equation}
\Gamma_\mu \equiv \frac{1}{2} \left[ u^\dagger (\partial_\mu-ir_\mu) u + u (\partial_\mu - il_\mu) u^\dagger \right]
\end{equation}
is the chiral connection. Under chiral transformations the covariant derivative transforms as
\begin{equation}
D_\mu \Psi \mapsto K(V_L,V_R,U)D_\mu \Psi.
\end{equation}
The LO pion-nucleon Lagrangian takes the form
\begin{equation}
\label{piNLag}
\mathscr{L}_{\pi N}^\text{LO} = \bar{\Psi}\left( i\slashed{D} - m_N +\frac{g_A}{2} \gamma^\mu\gamma_5 u_\mu\right) \Psi,
\end{equation}
where $m_N$ is the nucleon mass and $g_A$ the nucleon axial vector coupling, both in the chiral limit, and
\begin{equation}
\label{vielbein}
u_\mu \equiv i \left[ u^\dagger (\partial_\mu-ir_\mu) u - u (\partial_\mu - il_\mu) u^\dagger \right]
\end{equation}
is the chiral vielbein, which transforms as $u_\mu \mapsto Ku_\mu K^\dagger$.

Because the nucleon mass does not vanish in the chiral limit, (covariant, timelike) derivatives on the
nucleon field are not suppressed at low energies \cite{Gasser:1987rb}. The additional
building blocks in the nucleon sector count as
\begin{equation}
\Psi = {\cal O}(q^0), \quad D_\mu \Psi = {\cal O}(q^0), \quad u_\mu = {\cal O}(q)
\end{equation}
in the power counting scheme. However, the combination $(i\slashed{D} - m_N)\Psi$ is
suppressed at ${\cal O}(q)$.

\section{Quark Lorentz Violation and Local Symmetries}
\label{sec:quark-level}

We are now prepared to generalize the preceding analysis to deal with an action that contains
LV terms. The fundamental LV operators will be introduced for the quark fields, and we must
understand how these modify the building blocks that will be used to construct
Lagrangians at the hadronic level.
This will require us to generalize the external fields $r_{\mu}$ and $l_{\mu}$, and
any further quantities (such as $\Gamma_{\mu}$ or $u_{\mu}$) that depend on the external
fields will be modified by the Lorentz violation present in the theory.

In the present analysis, we restrict the discussion to the same LV dimension-4 quark operators as
we previously considered in~\cite{Kamand:2016xhv}; and again we only consider $u$ and $d$ quarks.
The resulting theory preserves CPT symmetry.
(Understanding the roles of dimension-3 mSME operators in hadronic physics remains a separate
outstanding question; many physical LV observables actually involve linear combinations of
dimension-3 and dimension-4 terms.) The Lagrange density corresponding to this two-flavor theory
can be written in the form\footnote{Also see the discussion in~\cite{Kamand:2016xhv}
regarding the reduction to two flavors. The three-flavor theory has been considered
in~\cite{ref-noordmans1}.}
\begin{equation}
\label{LV_lag}
\mathcal{L}_\text{light quarks}^{\text{CPT-even}}
= \frac{i}{2} \bar{Q}_{L} C_L^{\mu\nu} \gamma_{\mu} \overset{\leftrightarrow}{\mathcal{D}_\nu} Q_{L}
+ \frac{i}{2} \bar{Q}_{R} C_R^{\mu\nu}  \gamma_{\mu} \overset{\leftrightarrow}{\mathcal{D}_\nu} Q_{R},
\end{equation}
where the matrices
\begin{equation}
C_{L/R}^{\mu\nu}=\left[
\begin{array}{cc}
c^{\mu\nu}_{u_{L/R}} & 0 \\
0 & c^{\mu\nu}_{d_{L/R}}
\end{array}
\right]
\end{equation}
collect the dimensionless coupling coefficients, and $\mathcal{D}_\nu Q_{L/R}$ is the Standard Model
covariant derivative, including the coupling to gluons and electroweak gauge fields. (To avoid confusion,
we denote this derivative by ${\cal D}_{\nu}$ to distinguish it from the hadronic $D_{\nu}$ from section~\ref{sec:chpt}.)
We again restrict the discussion to the parts of $C_{L/R}^{\mu\nu}$ that are symmetric in the
Lorentz indices $(\mu,\nu)$. It will also be convenient to consider the isosinglet and isovector
parts of the $C_{L/R}^{\mu\nu}$ separately. They are defined by
\begin{equation}
\CLRsing^{\mu\nu} = \frac{1}{2} \tr (C_{L/R}^{\mu\nu}) ,\quad \CLRtrip^{\mu\nu}=C^{\mu\nu}_{L/R}-\mathds{1}\CLRsing^{\mu\nu},
\end{equation}
where $\mathds{1}$ is the identity in flavor space.

The Lagrangian of eq.~\eqref{LV_lag} is not invariant under global chiral transformations
since the matrices $C_{L/R}^{\mu\nu}$ are constant and do not transform. However, following
the example of the chiral-symmetry-breaking quark masses in the QCD Lagrangian, invariance
under chiral transformations can be restored by requiring that the $C_{L/R}^{\mu\nu}$ transform as
\begin{equation}
C_L^{\mu\nu} \mapsto L C_L^{\mu\nu} L^\dagger, \quad C_R^{\mu\nu} \mapsto R C_R^{\mu\nu} R^\dagger,
\end{equation}
or equivalently
\begin{equation}
\begin{split}
\CLsing^{\mu\nu}  & \mapsto \CLsing^{\mu\nu}  , \quad \CLtrip^{\mu\nu}    \mapsto L\CLtrip^{\mu\nu} L^{\dagger} \\
\CRsing^{\mu\nu}  & \mapsto \CRsing^{\mu\nu}  ,  \quad  \CRtrip^{\mu\nu}  \mapsto R\CRtrip^{\mu\nu} R^{\dagger}.
\end{split}
\end{equation}
Using this hypothetical transformation behavior, the $C_{L/R}^{\mu\nu}$ can be used as building blocks
in the effective hadronic Lagrangian, and the pattern of chiral symmetry breaking is thereby
mapped from the quark to the hadronic level. The leading-order results for the pion and pion-nucleon
Lagrangians without the coupling to external fields were given in~\cite{Kamand:2016xhv}, but these were
constructed by requiring invariance only under global chiral transformations.

Working with the derivatives on the quarks fields in eq.~\eqref{LV_lag} requires special care when
considering the {\em local} chiral transformations of eq.~\eqref{local_transform} and the inclusion of external fields.
The transformation properties of the $C_{L/R}^{\mu\nu}$ coefficients may be straightforwardly extended to
cover the local case,
\begin{equation}
\label{CLR_localtrans}
C_L^{\mu\nu} \mapsto V_L C_L^{\mu\nu} V_L^\dagger, \quad C_R^{\mu\nu} \mapsto V_R C_R^{\mu\nu} V_R^\dagger.
\end{equation}
However, under local transformations, the derivatives in eq.~\eqref{LV_lag} result in terms that
contain the LV coefficients $C_{L/R}^{\mu\nu}$ in combination with derivatives of the transformation
matrices, $\partial_\nu V_{L/R}$. These terms appear to break the local invariance of the Lagrange density
and thus should require additional modifications. However, the $\mathcal{D}_\nu Q_{L/R}$ are
the full standard model covariant derivatives, including not only the coupling to gluons, but also
to electroweak gauge fields. The introduction of the $C_{L/R}^{\mu\nu}$ thus not only modifies the
strong interactions, but also the coupling of quarks to the external fields in a well-defined manner.
There are two ways to treat the local invariance, both leading to the same result.

The first
(and simpler) method is to explicitly modify the Lagrange density that describes the coupling to
external fields. Since the covariant derivative in eq.~\eqref{LV_lag} is the standard model derivative
that also appears in the LC sector, the LV coupling to external fields follows the same pattern as the
kinetic term. The Lagrange density from eq.~\eqref{LagExtLC} is modified by a LV term $\mathcal{L}_\text{external}$,
\begin{equation}
\mathscr{L}_\text{external} \to  \mathscr{L}_\text{external} + \mathcal{L}_\text{external},
\end{equation}
where
\begin{equation}
\mathcal{L}_\text{external} = \frac{1}{2} \bar{Q}_L \gamma_\mu \{ C_L^{\mu\nu},l_\nu \} Q_L
+\frac{1}{2} \bar{Q}_R \gamma_\mu \{ C_R^{\mu\nu},r_\nu \} Q_R,
\end{equation}
where $\{\cdot,\cdot\}$ denotes the anticommutator.
The coupling to external fields can be described by replacing $r_\mu,$ $l_\mu,$ and  $v_{\mu}^{(s)}$
by the same expressions as in the LC case. [See, e.g., eq.~\eqref{ExEMCouple} for the electromagnetic case.]
To first order in the LV coefficients, the complete Lagrange density is then invariant under
local chiral transformations, given the transformation properties of the external fields listed
in eq.~\eqref{ext_transform}. In particular, this means that the covariant derivatives on the
pion field matrix [given in eq.~\eqref{pion_cov}] and on the nucleons [given in eq.~\eqref{NucCovDer}]
remain unaltered.

In the alternative approach, one can also choose to leave the external Lagrange density
$\mathscr{L}_\text{external}$ unchanged. To describe the \emph{same} LV interactions with
electroweak gauge fields at the quark level as in the first approach, the external fields have to be replaced by
\begin{equation}
\label{mod_ext_fields}
r^\mu\to r^\mu + \frac{1}{2}\{ C_R^{\mu\nu},r_\nu \}, \quad l_\mu \to l_\mu + \frac{1}{2}\{ C_L^{\mu\nu},l_\nu \}.
\end{equation}
In this approach, the presence of the derivatives on the quarks fields in eq.~\eqref{LV_lag}
leads to nontrivial modifications of the behavior of the external fields under local chiral
transformations. To first order in the LV coefficients, the combined Lagrangians, eqs.~\eqref{QCD-external}
and \eqref{LV_lag}, remain invariant under local transformations $(V_L,V_R)\in SU(2)_L \times SU(2)_R$
as long as the external fields $l_\mu$ and $r_\mu$ transform as
\begin{align}
\label{mod_ext_transform}
l^\mu & \mapsto V_L l^\mu V_L^\dagger + i V_L \partial^\mu V_L^\dagger +\frac{i}{2}
\left[  V_L C_L^{\mu\nu} \partial_\nu V_L^\dagger - \partial_\nu V_L  C_L^{\mu\nu} V_L^\dagger \right]  , \notag \\
r^\mu & \mapsto V_R r^\mu V_R^\dagger + i V_R \partial^\mu V_R^\dagger  +\frac{i}{2}
\left[  V_R C_R^{\mu\nu} \partial_\nu V_R^\dagger - \partial_\nu V_R  C_R^{\mu\nu} V_R^\dagger \right] .
\end{align}
This also leads to modifications of the pion and nucleon covariant derivatives. In order to maintain the desired property
\begin{equation}
D^\mu_\text{LV} U \mapsto V_R (D^\mu_\text{LV} U ) V_L^{\dag}
\end{equation}
under local transformations, the pion covariant derivative in the presence of the Lorentz violation has to take the form
\begin{equation}
\label{pion_LV_cov}
D^\mu_\text{LV} U \equiv D^\mu U +\frac{i}{2} \{ C_R^{\mu\nu},r_\nu \} U - \frac{i}{2} U\{ C_L^{\mu\nu},l_\nu \}.
\end{equation}
Similarly, the nucleon covariant derivative has to be replaced by
\begin{equation}
\label{LVNucCovDer}
D^\mu_\text{LV} \Psi \equiv \left(\partial^\mu + \Gamma^\mu_\text{LV} - i v^{(s)\mu} \right)\Psi,
\end{equation}
where
\begin{equation}
\Gamma^\mu_\text{LV} \equiv \Gamma^\mu + \frac{i}{4}u^\dagger \{C_R^{\mu\nu},r_\nu\} u + \frac{i}{4}u \{C_L^{\mu\nu},l_\nu\} u^\dagger.
\end{equation}
Since the chiral vielbein $u^\mu$ of eq.~\eqref{vielbein} contains the external fields,
in this approach it also has to be modified in the presence of Lorentz violation to preserve the desired
transformation property. Up to first order in the LV coefficients, it takes the form
\begin{equation}
\label{LVvielbein}
u^\mu_\text{LV} = u^\mu -\frac{1}{2}u^\dagger \{C_R^{\mu\nu},r_\nu\} u
+ \frac{1}{2}u \{C_L^{\mu\nu},l_\nu\} u^\dagger.
\end{equation}
Inserting the expressions from eq.~\eqref{mod_ext_fields} into the modified covariant derivatives
and chiral vielbein, the additional LV terms cancel, and one obtains the same expressions for
the covariant derivatives as in the first approach. However, this second approach is obviously more
cumbersome, and so in what follows, we shall exclusively use the first approach.

\section{Lorentz-Violating Effective Lagrangians}
\label{sec:LV_ChPT}

To utilize the results of experimental Lorentz tests, performed with real particles, it is
obviously necessary to have a description of Lorentz violation in terms of the physical quanta
of the theory---which are composite hadrons. The effective hadronic Lagrange densities we need
are constructed by writing down all terms that are consistent with the symmetry properties of
the underlying theory. In addition to local chiral invariance, the transformation behavior of
the quark-level Lagrange density of eq.~\eqref{LV_lag} under parity (P) and charge conjugation (C)
provides constraints on the operators that may be present. We will restrict our discussion
to terms at the lowest chiral order, and our calculations are similarly valid only up to linear order
in the LV coefficients.

The available building blocks are the pion and nucleon fields, their corresponding covariant derivatives,
the LV coefficients $C_{L/R}^{\mu\nu}$, the external fields, and (in the nucleon sector only) the chiral vielbein.
As shown in~\cite{Kamand:2016xhv}, the LO LV Lagrange densities are of chiral order ${\cal O}(q^2)$
in the pion sector and ${\cal O}(q^0)$ in the nucleon sector.
Since $\chi$ is ${\cal O}(q^2)$, it can be ignored in the nucleon case, although in principle it could
contribute to the LO pion Lagrangian.  However, the Lorentz indices on the $C_{L/R}^{\mu\nu}$
would require the introduction of additional pion covariant derivatives, each counting as $O(q)$.
We can therefore neglect $\chi$ entirely in the construction of the LO Lagrange densities.
The field strength tensors $f_{L/R}^{\mu\nu}$ are also of second order, but they are antisymmetric
in their Lorentz indices. Since we take the $C_{L/R}^{\mu\nu}$ to be symmetric, the $f_{L/R}^{\mu\nu}$
can also only contribute beyond LO.
Instead of treating the external fields $l_\mu$ and $r_\mu$ individually, it is convenient to
introduce covariant derivatives of the LV coefficients $C_{L/R}^{\mu\nu}$. Given the transformation
behavior from eq.~\eqref{CLR_localtrans}, these take the form~\cite{Fearing:1994ga}
\begin{align}
D^\rho C_L^{\mu\nu} &\equiv \partial^\rho C_L^{\mu\nu} - i l^\rho C_L^{\mu\nu}+i C_L^{\mu\nu} l^\rho
= - i l^\rho C_L^{\mu\nu}+i C_L^{\mu\nu} l^\rho \\
D^\rho C_R^{\mu\nu} &\equiv \partial^\rho C_R^{\mu\nu} - i r^\rho C_R^{\mu\nu}+i C_R^{\mu\nu} r^\rho
= - i r^\rho C_R^{\mu\nu}+i C_R^{\mu\nu} r^\rho ,
\end{align}
where we have used that the $C_{L/R}^{\mu\nu}$ are spacetime constants.
This means that the covariant derivatives of the isosinglet parts of the LV coefficients vanish identically.
The advantage of introducing the covariant derivatives is that they satisfy a product rule \cite{Fearing:1994ga},
which can be used to construct
integration by parts arguments and to identify total derivatives, so as to reduce the number of terms
in the Lagrange density to a minimal set.

\subsection{Mesonic Lagrangian}

The building blocks and their assumed transformation properties can now be used to construct
Lagrange densities that are invariant under local chiral transformations and that match the
discrete symmetry properties of the quark-level Lagrangian.
It can be shown that in considering the transition to local chiral transformations, the only
change compared to the globally invariant Lagrange density in~\cite{Kamand:2016xhv} is the
replacement of the standard derivatives
acting on the pion fields with the covariant derivatives of eq.~\eqref{pion_cov}.
The LO LV Lagrange density in the pion sector that is invariant under local chiral transformations is given by
\begin{equation}
\label{LV_Lag_pi}
\mathcal{L}_{\pi}^\text{LO} = \beta^{(1)} \frac{F^2}{4} (\CRsing^{\mu\nu} + \CLsing^{\mu\nu})
\tr\left[ ( D_\mu U )^\dagger   D_\nu U \right].
\end{equation}
$\beta^{(1)}$ turns out to be the sole low energy constant (LEC) at LO in this sector.

While, in principle, a number of other structures involving the $C_{L/R}^{\mu\nu}$ and two pion
covariant derivatives that have the correct transformation behavior under P, C, and chiral symmetry
can be written down, these terms all can be shown to be linearly dependent using identity relations
that are also commonly used in the LC theory, such as
\begin{align}
\label{LC_rel1} D^\mu U U^\dagger & = -U D^\mu U^\dagger , \\
\label{LC_rel2} \tr(D^\mu U U^\dagger) & = 0 ,\\
\label{LC_rel3} \tr(D^\mu D^\nu U U^\dagger) &= - \tr( D^\nu U D^\mu U^\dagger) = \tr( U D^\mu D^\nu U^\dagger) .
\end{align}

For the isovector LV coefficients $\CLRtrip^{\mu\nu}$, the covariant derivatives $D^\rho C_{L/R}^{\mu\nu}$
have to be considered as additional building blocks. At the leading order we are considering here,
expressions involving the covariant derivatives of the LV coefficients can be shown to be equivalent
to terms with pion covariant derivatives, by a combination of integrating by parts and applying the
relations (\ref{LC_rel1}--\ref{LC_rel3}).
As a result, the only independent term that can be written down in principle is analogous
to eq.~\eqref{LV_Lag_pi}, but involving the isovector pieces $\CLRtrip^{\mu\nu}$.
The corresponding term in the case without external fields was shown to
vanish in~\cite{Kamand:2016xhv}. Using trace relations that can be derived from the Cayley-Hamilton
theorem (see, e.g., the discussion in Ref.~\cite{Ebertshauser:2001nj}), the Lagrangian invariant under
local transformations and involving the isovector coefficients can still be shown to vanish, as we do in the appendix.
In spite of this, however, the covariant derivatives $D^\rho C_{L/R}^{\mu\nu}$ will
presumably become important for the construction of mesonic Lagrange densities beyond LO.

\subsection{Pion-Nucleon Lagrangian}

In analogy to the pion case, the LO LV pion-nucleon Lagrange density can be obtained
from the terms in~\cite{Kamand:2016xhv} by replacing each partial derivative $\partial^{\mu}\Psi$
on the nucleon field with the covariant derivative $D^\mu \Psi$.
As in the mesonic sector, one of the main tasks is the reduction of all possible locally
chiral invariant terms at a given order to a minimal set.
In the nucleon sector it is convenient to introduce the building blocks
\begin{equation}
\label{nuc_LV_coeff}
\tCRtrip^{\mu\nu} \equiv u^\dagger \CRtrip^{\mu\nu} u , \quad \tCLtrip^{\mu\nu} \equiv u \CLtrip^{\mu\nu} u^\dagger,
\end{equation}
which transform as
\begin{equation}
\tCLRtrip^{\mu\nu} \mapsto K \tCLRtrip^{\mu\nu} K^\dagger,
\end{equation}
with $K(V_L,V_R,U)$ defined in eq.~\eqref{KDef}. Eliminating
terms at LO requires the application of integration by parts and total derivative arguments.
For building blocks transforming like $\tCLRtrip^{\mu\nu}$, the covariant derivative is given
by~\cite{Bijnens:1999sh}
\begin{equation}
D_\rho \tCLRtrip^{\mu\nu} = \partial_\rho \tCLRtrip^{\mu\nu} + [\Gamma_\rho, \tCLRtrip^{\mu\nu}].
\end{equation}
The partial derivative on $\tCLRtrip^{\mu\nu}$ does not vanish, because of the presence of the
pion field matrices $u$ and $u^\dagger$. However, because $\partial_\rho u$, $\partial_\rho u^\dagger$,
$l_\rho$, and $r_\rho$ are all of order ${\cal O}(q)$, $D_\rho \tCLRtrip^{\mu\nu}$ is also of ${\cal O}(q)$.
We can therefore neglect all terms containing $D_\rho \tCLRtrip^{\mu\nu}$ at the LO to which we are working.
The minimal set of terms is given by those listed in~\cite{Kamand:2016xhv} using the covariant nucleon derivative,
\begin{equation}
\label{LVpiNLag}
\begin{split}
\mathcal{L}_{\pi N}^\text{LO}  = & \Big\{
\alpha^{(1)}\bar{\Psi}[(\tCRtrip^{\mu\nu} + \tCLtrip^{\mu\nu}) (\gamma_{\nu} i D_{\mu} + \gamma_{\mu} i D_{\nu})]\Psi \\
 & + \alpha^{(2)}\left({\CRsing^{\mu\nu}} + \CLsing^{\mu\nu}\right)\bar{\Psi}(\gamma_{\nu} i D_{\mu} + \gamma_{\mu}i D_{\nu})]\Psi  \\
 & + \alpha^{(3)}\bar{\Psi}[(\tCRtrip^{\mu\nu} - \tCLtrip^{\mu\nu}) (\gamma_{\nu}\gamma^{5} i D_{\mu} + \gamma_{\mu}\gamma^{5} i
 D_{\nu})]\Psi  \\
 & + \alpha^{(4)}\left(\CRsing^{\mu\nu} - \CLsing^{\mu\nu}\right) \bar{\Psi} (\gamma_{\nu}\gamma^{5} i D_{\mu} +
 \gamma_{\mu}\gamma^{5} i D_{\nu})\Psi\Big\} + h.c.
\end{split}
\end{equation}
The dimensionless LECs $\alpha^{(i)}$ are the same as the ones in~\cite{Kamand:2016xhv}.
It is a consequence of gauge invariance that the same LECs appear in the free nucleon sector
and in the nucleons' couplings to gauge fields. An important implication of this is that
LV observables derived entirely from the observation of free particle propagation may still be used to place direct
constraints on LV interaction coefficients.

\subsection{Couplings to Photons}

As an important representative example, we consider the pion-photon and nucleon-photon couplings.
Using the external fields from eq.~\eqref{ExEMCouple},
\begin{equation}
l_\mu=r_\mu= -\frac{e}{2} \mathcal{A}_\mu \tau_3,\quad v_\mu^{(s)} = -\frac{e}{2} \mathcal{A}_\mu ,
\end{equation}
the LO LV coupling of a pion to an electromagnetic potential is given by the Lagrange density
\begin{equation}
\mathcal{L}^\text{LO}_{\pi\gamma} = -e \beta^{(1)} \frac{1}{2}\left(c_{u_R}^{\mu\nu} + c_{d_R}^{\mu\nu}
+ c_{u_L}^{\mu\nu} + c_{d_L}^{\mu\nu}\right)\epsilon_{3ab}\mathcal{A}_\mu \phi^a \partial_\nu \phi^b,
\end{equation}
where $\phi^a$ is a Cartesian pion field with isospin index $a$. As expected from the form of the
Lagrange density $\mathcal{L}_{\pi}^\text{LO}$ from eq.~\eqref{LV_Lag_pi}, this is a straightforward
LV generalization of the standard coupling of a photon to a pion. The gauge potential ${\cal A}_{\mu}$
couples to a current that is modified by the Lorentz violation, and, in fact, this is exactly the
deformed current that would be expected from the pions' modified, LV kinetic term.
In the pure pion propagation and pion-photon coupling sectors, the modifications depend on a symmetric
linear combination of the dimension-4 LV coefficients we have been considering. This makes sense,
because in the chiral limit, each charged pion field contains equal contribution from the up- and
down-flavored, left- and right-chiral quark fields.

The coupling to the nucleon is described by the terms
\begin{align}
\label{eq:LNgamma}
\mathcal{L}^\text{LO}_{N\gamma} & = -4e \left[ \alpha^{(1)} (c_{u_R}^{\mu\nu}-c_{d_R}^{\mu\nu}+c_{u_L}^{\mu\nu}-c_{d_L}^{\mu\nu})
+\alpha^{(2)} (c_{u_R}^{\mu\nu}+c_{d_R}^{\mu\nu}+c_{u_L}^{\mu\nu}+c_{d_L}^{\mu\nu}) \right]
\bar{\Psi} \frac{1}{2}(\Id+\tau_3)\gamma_\mu \mathcal{A}_\nu \Psi \nonumber\\
&-4e \left[ \alpha^{(3)} (c_{u_R}^{\mu\nu}-c_{d_R}^{\mu\nu}-c_{u_L}^{\mu\nu}+c_{d_L}^{\mu\nu})
+\alpha^{(4)} (c_{u_R}^{\mu\nu}+c_{d_R}^{\mu\nu}-c_{u_L}^{\mu\nu}-c_{d_L}^{\mu\nu}) \right]
\bar{\Psi} \frac{1}{2}(\Id+\tau_3)\gamma_\mu \gamma_5\mathcal{A}_\nu \Psi.
\end{align}
Again, this appears to be a straightforward generalization of the pure nucleon kinetic terms. It is immediately
clear from the presence of $(\Id+\tau_3)$ that there are, at this order, no anomalous electromagnetic
couplings to the neutron. The proton is endowed with effective $c^{\mu\nu}$ and $d^{\mu\nu}$ terms of the form
\begin{equation}
{\cal L}_{p}=\bar{p}\left(\gamma^{\mu}+c_{p}^{\nu\mu}\gamma_{\nu}+d_{p}^{\nu\mu}\gamma_{5}\gamma_{\nu}\right)
\left(i\partial_{\mu}-e{\cal A}_{\mu}\right)p.
\end{equation}
This is just the form used in most phenomenalistic treatments of the SME's dimension-4 proton sector.
Any other combinations of the $C^{\mu\nu}_{L/R}$ parameters different from those given in eqs.~\eqref{LVpiNLag}
and \eqref{eq:LNgamma} cannot be separately observed through any LO nucleon propagation observables in \chiPT.

\subsection{Effects on $\beta$-Decay}

The generalization of the coupling of the nucleons to other boson fields are straightforward. In our
\chiPT framework, these couplings come in two principal types. The presence of the $\tCRtrip^{\mu\nu}$
and $\tCLtrip^{\mu\nu}$ terms in ${\cal L}_{\pi N}^{{\rm LO}}$ introduces couplings of the nucleons to
pions, because of the inclusion of the pion fields $u$ and $u^{\dag}$. The other couplings are to the remaining
massive vector bosons in the electroweak sector, which appear as straightforward generalizations of the
photon coupling. Such terms would introduce, for example, Lorentz violation in the matrix element for
neutron $\beta$-decay.

Up to this point, only the simpler influence of pure weak-sector Lorentz violation on that
matrix element has been studied~\cite{Noordmans:2013xga,Noordmans:2013dha}. It turns out that the Lorentz violation
in the quark sector has similar effects on nuclear $\beta$-decays, but the specific coefficients involved depend on the
precise nature of the decay process (whereas the effects of weak-sector Lorentz violation are more universal, with
the same parameters affecting all types of decays).

In order to have a consistent theory for W$^{\pm}$-mediated processes, we must enforce the $SU(2)_{L}$ gauge symmetry. There
cannot be different coefficients in the kinetic terms for the left-handed up and down quarks. This means that
$^{3}C_{L}^{\mu\nu}=0$ (or $c_{u_{L}}^{\mu\nu}=c_{d_{L}}^{\mu\nu}=c_{L}^{\mu\nu}$),
reducing the number of independent coefficients by one fourth. Then the weak coupling may be
included in the Lagrange density by taking
\begin{equation}
l_{\mu}=-\frac{gV_{ud}}{\sqrt{2}}\left({\cal W}_{\mu}^{+}\tau^{+}+{\cal W}_{\mu}^{-}\tau^{-}\right).
\end{equation}
With this external field,
the Lagrange density for the nucleonic weak interaction vertex takes a form very similar to the nucleon-photon vertex
in eq.~\eqref{eq:LNgamma}. The only additional complexity comes from the presence of the matrix $\tau^{3}$ in
$^{3}C_{L}^{\mu\nu}$, which changes some signs via $\tau^{3}\left({\cal W}_{\mu}^{+}\tau^{+}+{\cal W}_{\mu}^{-}\tau^{-}\right)=
{\cal W}_{\mu}^{+}\tau^{+}-{\cal W}_{\mu}^{-}\tau^{-}$.

In the effective action for low-energy $\beta$-decay (neglecting the momentum dependence of the W$^{\pm}$ propagator),
the electroweak $\chi^{\mu\nu}$ coefficients that were studied in~\cite{Noordmans:2013xga,Noordmans:2013dha} enter in the form
\begin{equation}
\mathcal{L}_{\beta}=-\chi^{\mu\rho}\left[\bar{p}\gamma_{\mu}\left(C_{V}+C_{A}\gamma_{5}\right)n\right]
\left[\bar{e}\gamma_{\rho}\left(1-\gamma_{5}\right)\nu\right]+h.c.,
\end{equation}
in terms of the proton, neutron, electron, and neutrino fields; the constants $C_{V}$ and $C_{A}$ describe the usual vector
and axial vector couplings of the nucleon, incorporating both the Fermi coupling and QCD matrix elements. 

However, the hadronic coefficients operate more
generally than the electroweak ones $\chi^{\mu\nu}_{EW}$---in two separate ways. There are different $\chi^{\mu\nu}$ tensors
for Fermi processes, $\chi_{F\pm}^{\mu\nu}$, and
Gamow-Teller processes, $\chi_{GT\pm}^{\mu\nu}$. They are given by
\begin{eqnarray}
\chi_{F\pm}^{\mu\nu} & = & \chi_{EW}^{\mu\nu}\pm\alpha^{(1)}\left(c_{u_{R}}^{\mu\nu}-c_{d_{R}}^{\mu\nu}\right)
+\alpha^{(2)}\left(2c_{L}^{\mu\nu}+c_{u_{R}}^{\mu\nu}+c_{d_{R}}^{\mu\nu}\right) \\
\chi_{GT\pm}^{\mu\nu} & = & \chi_{EW}^{\mu\nu}\pm\alpha^{(3)}\left(c_{u_{R}}^{\mu\nu}-c_{d_{R}}^{\mu\nu}\right)
-\alpha^{(4)}\left(2c_{L}^{\mu\nu}-c_{u_{R}}^{\mu\nu}-c_{d_{R}}^{\mu\nu}\right).
\end{eqnarray}
The $\pm$ indices denote the sign of the intermediate vector boson that mediates the decay: $-$ for electron emission; $+$ for electron
capture or positron emission. The four types of decays therefore probe four different linear combinations of parameters.

It may initially seem puzzling that the linear combinations of coefficients that appear include parameters for 
both left- and also right-handed quarks. Since the W$^{\pm}$ bosons couple, at the fundamental level, only to left-chiral fields,
having contributions from $C_{R}^{\mu\nu}$ comparable to those from $C_{L}^{\mu\nu}$ may not be expected.
However, it is actually reasonable for both sets of coefficients to be involved. The gauge couplings involve matrix elements of
momentum or angular momentum operators between the proton and neutron fields. These matrix elements are modified by the Lorentz
violation. The hadronic fields being thus probed are composites, with
the constituent partons each carrying only a fraction of a nucleon's momentum (or spin). The W$^{\pm}$ couples directly to
the a left-chiral valance quark, and so the underlying coupling depends on the momentum carried by specifically that
valance quark. Yet the momentum and spin available to be carried by the valance quark that interacts with the boson depend on how
much is being carried by the other valance quarks---at least one of which is necessarily right handed. Thus the momentum and
angular momentum matrix elements of the right-chiral fields, including the modifications due to Lorentz violation, do play a role in
the effective W$^{\pm}$ couplings. This would be the case even in the absence of the Lorentz violation, but without having the
different $C_{L}^{\mu\nu}$ and $C_{R}^{\mu\nu}$, the indirect dependence on the right-chiral sector is effectively invisible.

In fact, the presence of the right-handed coefficients actually provides some guidance as to the relative signs and
magnitudes of the LECs.
In the decay of a left-chiral neutron, for example, the W$^{-}$ involved couples directly to a left-chiral down quark. The two
remaining valance quarks are essentially equal combinations of up and down flavor, left and right chirality. In order that the
relevant linear combination of $\chi^{\mu\nu}$ parameters involved in the decay matrix element,
$C_{V}\chi_{F-}^{\mu\nu}-C_{A}\chi_{GT-}^{\mu\nu}$, not be controlled mostly by $C_{R}^{\mu\nu}$, the LECs $\alpha^{(2)}$
and $\alpha^{(4)}$ should have the same sign. Also, the coefficients $\alpha^{(1)}$ and $\alpha^{(3)}$ should not dominate
over $\alpha^{(2)}$ and $\alpha^{(4)}$.

The most recent direct search for anisotropy in $\beta$-decay examined the lifetime of $^{20}$Na, decaying via a Gamow-Teller
positron emission, looking at how that lifetime depended on the spin of the $^{20}$Na nucleus, all while the
terrestrial laboratory was rotating~\cite{ref-sytema}. The experiment had sensitivity at the $4\times10^{-4}$ level to
the imaginary parts of the coefficients $\chi_{EW}^{\mu\nu}$ (which are, unlike the quark coefficients, generally complex).
This kind of $\beta$-decay measurement is not necessarily directly competitive with the much stronger bounds on proton and neutron
coefficients in atomic experiments, but they do provide potential access to different linear combinations.

In fact, the much tighter
bounds on the proton and neutron kinetic coefficients actually simplify an analysis of the effects of the quark
$\chi^{\mu\nu}$ contributions substantially. In a general theory, the hadronic coefficients would have purely kinematic effects on
the decay rate---changing the available phase space by shifting the energies of the initial and final nuclear states. However,
for decay-based experiments with sensitivities at the $\sim10^{-4}$ level, these kinematic factors may be neglected, because
the linear combinations of coefficients that contribute to nuclear energy shifts are already so well constrained. In effect, the
known nucleon propagator constraints mean that the primary sensitivity of such a $\beta$-decay experiment is to the $\chi^{\mu\nu}_{EW}$
parameters, which can only be measured via such decay processes.

There is a similar analysis possible for the weak decays of charged mesons, particularly the charged pions. Once again, there
is an effective $\chi^{\mu\nu}$ parameter for the decay.  At tree level, where the
$\langle{\cal W}^{+}_{\mu}{\cal W}^{-}_{\nu}\rangle$ propagator is effectively just
$-i\left(g_{\mu\nu}+\chi_{EW}^{\mu\nu})\right/M_{W}^{2}$, forms of Lorentz violation in the lepton, electroweak, and quark sectors
enter indistinguishably in the matrix element for a process such as $\pi^{-}\rightarrow \mu^{-}+\bar{\nu}_{\mu}$.
The general structure of the matrix element for the process, in terms of the pion momentum $p^{\pi}$, is
\begin{equation}
\label{eq-Mdecay}
{\cal M}\propto\left[\bar{u}_{\ell}\left(g^{\mu\nu}+c^{\mu\nu}_{\ell_{L}}\right)\gamma_{\mu}v_{\nu_{\ell}}\right]
\left[g_{\nu\rho}+\left(\chi_{EW}\right)_{\nu\rho}\right]
\left[g^{\rho\sigma}+\beta^{(1)}c^{\rho\sigma}_{L}\right]p_{\sigma}^{\pi}.
\end{equation}
The $\bar{u}_{\ell}$ and $v_{\nu_{\ell}}$ are spinors for the charged lepton and neutrino involved, and the
$c^{\mu\nu}_{\ell_{L}}$ are coefficients for Lorentz violation in the lepton sector. The three bracketed terms in
eq.~\eqref{eq-Mdecay} come from the lepton vertex, vector boson propagator, and pion vertex, respectively.
So at first order, the net Lorentz-violating contributions to the matrix element depend only on the linear combination
\begin{equation}
\chi_{\pi}^{\mu\nu}=\chi_{EW}^{\mu\nu}+c_{\ell_{L}}^{\mu\nu}+\beta^{(1)}c^{\mu\nu}_{L}.
\end{equation}
In this case, it is true that only the $c^{\mu\nu}_{L}$ quark coefficients contribute. The reason is that the amplitude for the
weak $\pi^{-}$ decay comes entirely from the portion of the pion wave function with valance left-chiral down quark and
right-chiral up antiquark.

Data collected in a search for temporal oscillations in the MINOS near detector neutrino and antineutrino
signals~\cite{ref-adamson1,ref-adamson3} has previously been interpreted as constraining the magnitude of the possible Lorentz
violation in the pion decay matrix element. There have also been direct studies of whether the lifetimes of moving pions obey the
standard time dilation formula~\cite{ref-nielsen}. Successive reanalyses have considered the impact of the second-generation lepton
coefficients $c_{\ell_{L}}^{\mu\nu}$~\cite{ref-altschul33}, the electroweak $\chi^{\mu\nu}_{EW}$~\cite{ref-altschul34}, and the quark
parameters in a simplified model without using the full apparatus of $\chi$PT~\cite{ref-noordmans2}. The present work provides a
rigorous description of how the Lorentz violation coefficients in the pion kinetic Lagrangian relate to those appearing in the decay
matrix element. The overall sensitivities on the linear combinations $\chi_{\pi}^{\mu\nu}$ are at the $\sim 10^{-3}$--$10^{-5}$
level, with the weakest constraints on the $\chi_{\pi}^{TT}$ coefficients that control isotropic violation of boost invariance.
Again, however, bounds based on observations of the nucleon kinetic coefficients and couplings to photons constrain the
$\beta^{(1)}c^{\mu\nu}_{L}$ much more strongly than either observations of pion kinematics or charged pion decays. So the primary
practical sensitivity of Lorentz tests involving $\pi^{-}\rightarrow \mu^{-}+\bar{\nu}_{\mu}$ is to the
$\chi_{EW}^{\mu\nu}+c_{\ell_{L}}^{\mu\nu}$ linear combination.

\section{Conclusions and Outlook}
\label{sec:concl}

We have extended our earlier \chiPT analysis to include local chiral transformations. The most important
product of this analysis is that we have derived definite forms for the LO couplings of pions and nucleons
to external gauge fields. This provides a rigorous justification for the way proton Lorentz violation has
been handled in many previous analyses, but it also demonstrates that the conclusions of some existing
work are not well justified. Moreover, we have also determined the LO LV couplings between pions and nucleons,
generalizing the usual LC interaction terms.

Now that it is known how LV modifies the couplings between hadrons and photons, it is possible to reexamine
the results of previous studies, to see whether the conclusions of earlier analyses were justified.
One of the obvious consequences of terms like $c_{p}^{\nu\mu}$ and $d_{p}^{\nu\mu}$ is that sufficiently
energetic charged particles endowed with such coefficients may have speeds exceeding 1. This will
result in the emission of vacuum Cerenkov photons~\cite{ref-altschulXXX}, analogous to ordinary
Cerenkov radiation in matter. In matter, the phase speed of light is typically diminished, and it may
be surpassed by sufficiently relativistic charges; in a LV theory, not all particles have the same
maximum achievable velocities in vacuum, and it may simply be possible for the charges to outpace
electromagnetic waves, thereby producing radiation.

There has recently been a claim that the absence of significant vacuum Cerenkov emission by hadronic
cosmic ray primaries can be used to place bounds on a number of quark-level SME
parameters~\cite{ref-schreck1}. Particles above the Cerenkov threshold would lose energy rapidly until
they eventually dropped below the threshold. The Cerenkov emission would be a very efficient energy
loss mechanism, and particles above the threshold energy  would not be able to traverse significant astronomical
distances without bleeding off much of their energies.
The observation of a cosmic ray at a given energy thus indicates that the observed energy
must be below the threshold, which translates directly into a constraint on a direction-dependent
combination of SME parameters.

This technique for placing constraints on SME parameters has been applied uncontroversially to get
bounds on the effective SME parameters for hadrons, primarily protons. However, the recent
claim~\cite{ref-schreck1} that the same method may be used to bound the underlying quark
parameters seems slightly problematic. Quarks are confined in hadrons, and the emission of electromagnetic
radiation by a color-neutral bound state is produced by a coherent combination of the radiation
from the constituent quarks. The radiation cannot so easily be resolved into contributions from individual
quarks.

This is borne out by our analysis in this paper. The hadrons couple to the electromagnetic field only through
certain combinations of quark coefficients. At LO, those specific linear combinations are the ones listed
in eq.~\eqref{eq:LNgamma}. The inclusion of additional terms from \chiPT beyond LO would modify the particular linear
combinations and make those combinations momentum dependent.
Deviations from the forms dictated by \chiPT come into play once the momentum transfer between photons and nucleons
becomes large enough such that the \chiPT expansion breaks down.
For a nucleon at rest, that requires a momentum transfer of $\sim350$ MeV, which corresponds to a substantial fraction of the
rest energy of the nucleon. The Cerenkov-emitting hadrons discussed in~\cite{ref-schreck1} are, of course,
not at rest; however, for approximately collinear reactions involving only ultrarelativistic quanta, the
fractions of the total momentum carried off by each of the outgoing particles is nearly independent of
the frame in which the reaction process is analyzed.
At the absolute Cerenkov threshold, only the emission of very soft Cerenkov photons is allowed, and therefore
the linear combination of LV coefficients in eq.~\eqref{eq:LNgamma} should be dominant.

In order to disturb the internal structure of
a radiating proton sufficiently to get an effective coupling to an individual quark's LV coefficients
(rather than the \chiPT coefficients determined for the full proton) a Cerenkov photon must carry away
a substantial fraction of the proton's total momentum.
However, the threshold for an
emission event in which the proton looses a fraction $\zeta$ of its momentum is larger by a factor of
$(1-\zeta)^{-1/2}$ than the threshold for soft emissions. Probing individual LV quark couplings requires
the emission of photons with large energies, for which the thresholds are
correspondingly greater. We conclude that, at a minimum, the quark bounds quoted in~\cite{ref-schreck1}
cannot be considered quantitatively reliable, since they do not take into account the higher energies
needed to probe beyond the \chiPT limit. With our current level of understanding, it is only possible to
use the absence of vacuum Cerenkov emission from primary cosmic rays to place robust bounds on
the kinetic coefficients for the composite hadrons involved, not on arbitrary dimension-4 LV quark parameters.

This conclusion makes a great deal of sense. The rate of a process such as vacuum Cerenkov radiation
is primarily determined by phase space availability. The fact that the process is forbidden by energy-momentum
conservation in the usual LC theory is equivalent to there being zero phase space available for the decay products.
Conversely, the process may be allowed to occur above a sufficient threshold energy in the LV theory, because with
modified dispersion relations for the quanta, there is phase space available for the process.
Phase space availability is a property of asymptotic states, but the up and down quarks do not exist as
asymptotic states. The quanta which have well-defined energy-momentum relations are the hadronic excitations
of the theory---which are pions and nucleons in our two-flavor theory.

Other past work has explored the influence of nucleons' motions inside large nuclei and how that motion affects
precisely which LV coefficients are measurable in atomic clock experiments~\cite{ref-altschulYYY}. To the extent that
the short-distance nucleon-nucleon potential responsible for nuclear binding is mediated by the exchange
of multiple pions, the LV interaction Lagrangians described in this work may be used to understand how Lorentz
violation would modify the shapes of the nucleon-nucleon potentials. This will enable a more precise
disentanglement of the various coefficients in the SME.

Moreover, while the present analysis was restricted to the dimension-4 coefficients, similar arguments should also restrict
which dimension-3 mSME coefficients in the nucleon sector can be constrained. Further generalizing the LV version
of \chiPT to address the behavior of the dimension-3 terms would be of significant interest. There are also
terms that appear beyond LO, such as LV modifications of hadrons' anomalous magnetic moment terms, which may be
significant in precision atomic clock tests of Lorentz symmetry. The couplings to the massive vector boson Z$^{0}$ will
also produce potentially interesting new LV observables. In general, \chiPT will continue to be a useful tool for
understanding the strongly interacting sector of the SME.

\section*{Acknowledgments}
This material is based upon work supported by the U.S. Department of Energy, Office of Science, Office of
Nuclear Physics, under Award Number DE-SC0010300 (RK and MRS).

\appendix

\section*{Appendix: Isovector Mesonic Sector}
\label{app-isovector}

The most general pion Lagrangian involving the isovector portions of the LV
coefficients $\CLRtrip$ can be shown to consist of a single independent term analogous to that of eq.~\eqref{LV_Lag_pi},
\begin{equation}
\label{LV_pi-isovector}
\frac{F^2}{4} \tr\left[ ( D_\mu U )^\dagger  \CRtrip^{\mu\nu} D_\nu U + D_\mu U \CLtrip^{\mu\nu} D_\nu U^\dagger \right ].
\end{equation}
To show that this term vanishes identically, it is convenient to rewrite it using the building
blocks from the nucleon sector, in particular the chiral vielbein of eq.~\eqref{vielbein} and the LV
coefficients in the form of eq.~\eqref{nuc_LV_coeff}.
It can be shown that eq.~\eqref{LV_pi-isovector} is proportional to
\begin{equation}
\tr \left[ u_\mu u_\nu (\tCRtrip^{\mu\nu}+\tCLtrip^{\mu\nu})  \right] =
\frac{1}{2} \tr \left[ \{u_\mu , u_\nu \} (\tCRtrip^{\mu\nu}+\tCLtrip^{\mu\nu})  \right] ,
\end{equation}
where we have used that we only consider the symmetric part of the LV coefficients.
The Cayley-Hamilton theorem for $2\times2$ matrices $A$ and $B$ implies
\begin{equation}
\{A,B\} = \tr(A) B + \tr(B) A + \tr(A B) \Id - \tr(A)\tr(B) \Id .
\end{equation}
For traceless matrices this reduces to
\begin{equation}
\{ A,B \} = \tr(AB) \Id ,
\end{equation}
which in turn implies
\begin{equation}
\tr ( \{ A,B \} C ) = \tr(AB) \tr(C)
\end{equation}
for any $2\times2$ matrix $C$; in particular, if $\tr(C) = 0$, then
\begin{equation}
\tr ( \{ A,B \} C ) = 0.
\end{equation}
Since $u_\mu$ and the $\tCLRtrip$ are all traceless, it follows that the expression in eq.~\eqref{LV_pi-isovector}
is identically zero, and so the isovector parts of the LV coefficients do not contribute at LO in the pion sector.


\end{document}